\documentclass[prl,amsmath,amssymb,twocolumn, showpacs, superscriptaddress,10pt]{revtex4}

\usepackage{amsmath}
\usepackage{hyperref}
\usepackage{graphicx}
\usepackage{amsfonts}
\usepackage{amsthm}
\usepackage{cases}
\usepackage{changepage}
\usepackage{placeins}
\usepackage{bm}
\usepackage{ulem}

\usepackage{color}
\definecolor{Blue}{rgb}{0.00, 0.00, 1.00}
\definecolor{Red}{rgb}{1.00, 0.00, 0.00}

\hypersetup{
    colorlinks=true,       
    linkcolor=red,          
    citecolor=blue,        
    filecolor=magenta,      
    urlcolor=cyan           
}

\newcommand{\be}{\begin{equation}}
\newcommand{\ee}{\end{equation}}
\newcommand{\bea}{\begin{eqnarray}}
\newcommand{\eea}{\end{eqnarray}}
\newcommand{\ba}{\begin{align}}
\newcommand{\ea}{\end{align}}

\allowdisplaybreaks

\usepackage{graphicx}
\begin{document}

\title{Spatial clustering of depinning avalanches in presence of long-range interactions}

\author{Cl{\'e}ment Le Priol}
\affiliation{Laboratoire de Physique de l'Ecole Normale Sup\'erieure, Universit\'e PSL, CNRS, Sorbonne Universit\'e, Universit\'e de Paris, 24 rue Lhomond, 75231 Paris Cedex, France}
\author{Pierre Le Doussal}
\affiliation{Laboratoire de Physique de l'Ecole Normale Sup\'erieure, Universit\'e PSL, CNRS, Sorbonne Universit\'e, Universit\'e de Paris, 24 rue Lhomond, 75231 Paris Cedex, France}
\author{Alberto Rosso}
\affiliation{LPTMS, CNRS, Universit\'e Paris-Sud, Universit\'e Paris-Saclay, 91405 Orsay, France}

\date{\today}

\begin{abstract}
Disordered elastic interfaces display avalanche dynamics at the depinning transition. For short-range  interactions, avalanches correspond to compact reorganizations of the interface well described by the depinning theory. For long-range elasticity, an avalanche is a collection of spatially disconnected clusters. In this paper we  determine the scaling properties of the clusters and relate them to the roughness exponent of the interface. The key observation of our analysis is the identification of a Bienaym{\'e}-Galton-Watson process describing the statistics of the cluster number. Our work has a concrete importance for experimental applications where the cluster statistics is a key probe of avalanche dynamics.
\end{abstract}


\maketitle

Many catastrophic phenomena such as epidemic outbreaks, earthquakes~\cite{fisher1997, jagla2014, dearcangelis2016} or financial crashes are initiated by a single unstable (or infected) seed that destabilizes many other elements. 
The instability propagates as a cascade where each unstable element triggers a number of offsprings.
These processes are also observed in the response of disordered systems to small perturbations and are called avalanches~\cite{sethna2001}. 
The cascades of plastic events in amorphous materials~\cite{lin2014, baret2002}, 
the ground-state reorganizations of mean-field spin glasses~\cite{ledoussal2012a, franz2017} 
or the jerky motion of fronts propagating in heterogeneous media~\cite{fisher1998, kardar1998}
are examples of such avalanches.

We can consider that the first model of avalanches was
introduced by Bienaym{\'e}~\cite{bienayme1845} and by Galton and Watson~\cite{watson1875} who were interested in the extinction probability of surnames.
In this model the number of offsprings per individual is an independent identically distributed
random variable whose average $R_0$ determines the fate of the process. 
If $R_0 > 1$ there is a finite probability that the surname never gets extinct,
while for $R_0 \leq 1$ the surname gets extinct with probability one.
In this case the family size $S$ (total number of descendants of an individual)
is a stochastic variable 
displaying a power law distribution $P(S)\sim S^{-\tau}$, with exponent $\tau=3/2$, which is truncated (exponentially) above a maximal size $S_m \sim (1-R_0)^{-2}$.
  
\begin{figure}[t]
\includegraphics[width=1 \linewidth]{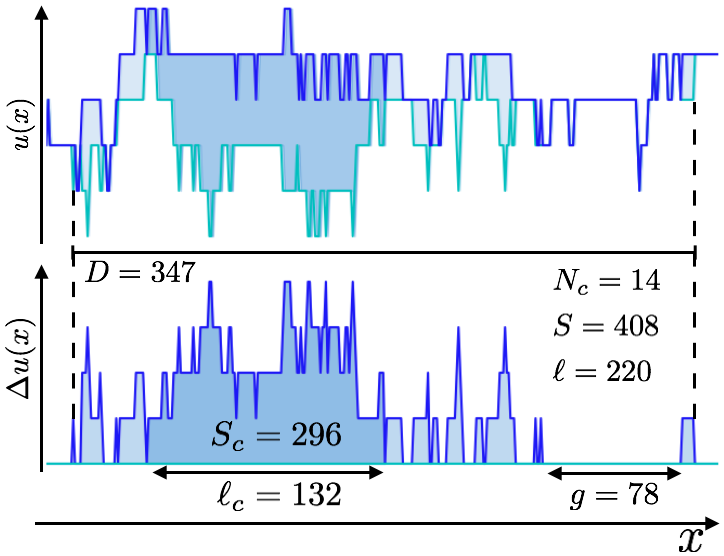}
\caption{Example of a LR avalanche for $\alpha=1$. \textbf{Top :} A one-dimensional front is pinned in a stable configuration at position $u(x)$ (light blue line). A small perturbation induces a large rearrangement of the front which reaches a new stable configuration (dark blue line). The area swept by the line is the avalanche size $S$.  
\textbf{Bottom :} Relative displacement $\Delta u(x)$ of the final configuration with respect to the initial one. The avalanche is made of $N_c$ distinct clusters whose statistical properties are the subject of this paper.
  \label{Fig:Intro}}
\end{figure}

We are interested in the avalanches observed when elastic interfaces of dimension $d$, such as magnetic domain walls~\cite{zapperi1998, laurson2013, durin2016}, crack fronts~\cite{gao1989, tanguy1998, bonamy2008, bonamy2011, ponson2016}
or wetting lines~\cite{joanny1984, roux2003, moulinet2004, ledoussal2009a}, propagate in heterogeneous materials.
Under the action of an external force $f$, the front remains pinned up to a critical force $f_c$ 
where it undergoes a depinning transition.
At the depinning point the front is in a random self-affine configuration 
characterized by the roughness exponent $\zeta$, whose value is known from numerical simulations~\cite{rosso2002, rosso2003} and Functional Renormalization Group calculations~\cite{nattermann1992a, narayan1993, ertas1994, ledoussal2002}.
Near depinning, small perturbations can trigger avalanches which are large rearrangements of the front, of total size $S$.
In presence of short-range (SR) elasticity, avalanches are spatially connected 
and the size exponent $\tau$ is smaller than the one predicted by the Bienaym{\'e}-Galton-Watson (BGW) model.
Indeed spatial information is absent from the BGW model while one expects that, 
for an unstable point of the interface, the analog of the number of offsprings depends on its generation and its position. 
At variance with BGW, the avalanche of an interface has a spatial location characterized by
a linear size $\ell$ that displays a power law distribution $P(\ell) \sim \ell^{-\kappa}$. 
In the framework of the depinning theory we know that $S\sim \ell^{d+\zeta}$ and that the avalanche exponents are not independent but related to the roughness exponent via the relations
$\tau = 2 - \frac{2}{d+\zeta}$ and  
$\kappa = d+\zeta-1$~\cite{narayan1993}.

In many physical situations, fronts are however characterized by a long-range (LR) elastic kernel that decays as $1/r^{d+\alpha}$.
In particular it has been shown that  the elasticity of crack fronts~\cite{rice1985, gao1989} and wetting lines~\cite{joanny1984} is LR with $\alpha=1$.
The shape of the interface is affected by the range of the interactions.
As a function of $\alpha$ we idenfity three regimes~:
(i) the mean-field (MF) case, for $\alpha \leq \frac{d}{2}$, where the interface is flat and $\zeta=0$,
(ii) for $\frac{d}{2} < \alpha < 2$ the interface is rough and $\zeta$ grows with $\alpha$,
(iii) for $\alpha \geq 2$, $\zeta$ saturates to its SR value.
LR interactions also affect the avalanche statistics, the depinning theory extends to $\alpha \leq 2$ the scaling relation for the size distribution. In particular in the MF case we recover the BGW value $\tau = 3/2$, while for $\frac{d}{2} \leq \alpha \leq 2$ we have~\cite{narayan1993, zapperi1998, dobrinevski2014}~:
\begin{align}
\tau &=  2- \frac{\alpha}{d+\zeta} \, , \label{eq: tau_NF} \\
\kappa &= 1 + d+\zeta -\alpha \, . \label{eq: kappa_NF}
\end{align}
The second equation derives from the first one via the self-affinity relation $S\sim \ell^{d+\zeta}$.
Beyond this modification, LR avalanches are qualitatively different from SR ones as they are in general disconnected objects.
An example of an avalanche for $\alpha=1$ is shown in Fig.~\ref{Fig:Intro}.
It is composed of $N_c$ disconnected clusters. We denote $S_c$ the (random) size of a given cluster,
and $\ell_c$ its extension. A gap between two clusters is denoted by $g$. 
We define the linear size of the avalanche as the sum of the extensions of its clusters,
noted schematically
$\ell = \sum \ell_c$. 
This length is much shorter than the diameter $D$ of the region affected by the reorganization.

All these quantities display power law statistics with exponents that we aim to determine. 
This is particularly important for experiments where the spatial structure is accessible~\cite{maloy2006, tallakstad2011}. In these
experiments the perturbation is provided by a continuous drive and independently triggered avalanches can overlap in time and/or space~\cite{bares2013, janicevic2016, lepriol2020}.
This makes the reconstruction of LR avalanches difficult and clusters, which are simple to identify, remain the reliable probe
of the universal properties of avalanche dynamics.

The first attempt to characterize the cluster statistics of LR depinning avalanches was done in Ref.~\cite{laurson2010},
where, based on numerical simulations for $\alpha=1$, a scaling relation for the cluster size exponent was proposed 
and justified by arguments that we show to be incorrect.
This relation was experimentally confirmed in~\cite{tallakstad2011} for $\alpha=1$.
In this Letter we derive scaling relations for all cluster exponents, and for all
$\frac{d}{2} \leq \alpha \leq 2$ and show that they depend only on the roughness exponent $\zeta$ 
(see table \ref{tab: exponents} for a summary of the results). 
The validity of our relations is based on
the key numerical observation, unveiled here, 
that the distribution of $N_c$ appears to obey the
BGW statistics : each cluster triggers a number of offsprings which can be considered as an independent and identically distributed random variable.

\begin{figure}[t!]
\includegraphics[width=1 \linewidth]{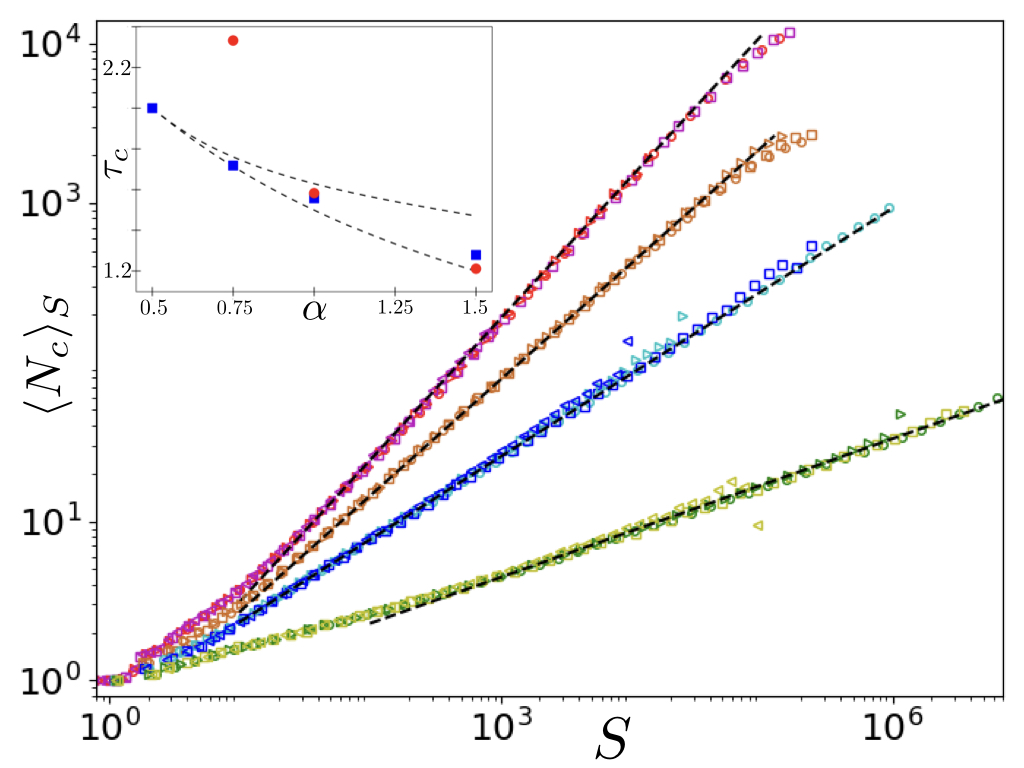}
\caption{\textbf{Main panel :} $\langle N_c \rangle_S$ versus $S$ for $\alpha=0.5$ (red and magenta), 0.75 (brown), 1 (blue and cyan) and 1.5 (green and yellow) (top to bottom). 
The dashed lines are fit of the exponent $\gamma_S$ with the values listed in table~\ref{tab: exponents}.
	\textbf{Inset :}  The values of $\gamma_S$ are fed into the prediction~\eqref{eq: Zapperi conj. generalized} (red circles) which can be compared with the direct measurement of $\tau_c$ (blue squares).
	The prediction is close from the measurement for $\alpha=1$ but becomes too large at low values of $\alpha$ (the prediction for $\alpha=0.5$ is $\tau_c=5.55$ and lies outside the figure).
	The dashed lines test our prediction of $\tau_c$ as a function of $\zeta$ (see table~\ref{tab: exponents}) with $\zeta^{1\text{loop}} = \frac{2\alpha-1}{3}$~\cite{ertas1994, ledoussal2002} (bottom line) and 
	$\zeta^{2\text{loop}} = \zeta^{1\text{loop}} + \frac{(2\alpha-1)^2}{3}\frac{\psi(1)+\psi(\alpha)-2\psi(\alpha/2)}{9\sqrt{2}\gamma}$ with $\psi$ the digamma function
and $\gamma \simeq 0.548$~\cite{ledoussal2002}~(upper line).
	\label{Fig: 2}}
\end{figure}

Let us start by defining the cellular automaton model, already used in~\cite{schmittbuhl1995, laurson2010, lepriol2020}, for the simulations of depinning interfaces.
Consider a one-dimensional interface (i.e. a line) $u(x,t)$ where $t$ is the time, $x$ the internal coordinate and $u$ the displacement field (see Fig.~\ref{Fig:Intro}). 
We assume that all time and space variables $t, x$ and $u$ are integers,
that the line size is $L$ and we implement periodic boundary conditions
$u(x+L,t)=u(x,t)$.
The disorder pins the line up to a local threshold force $\eta^{th}(x,u)$ balanced by the force 
\begin{equation}
F(x,t) = m^2 \left( w - u(x,t) \right) + \sum_{x'} \frac{u(x',t) - u(x,t)}{|x'-x|^{1+\alpha}} \, ,
\end{equation}
where the first term describes the driving force $f(x,t)=m^2 \left( w - u(x,t) \right)$ deriving from a harmonic confinement of curvature $m^2$ and the second is the LR elastic force.
The local thresholds $\eta^{th}(x,u)$ are independent identically distributed numbers drawn from the positive part of a normal distribution. 
The harmonic confinement induces a characteristic transversal length $\ell_m \sim m^{-2/\alpha}$ beyond which the interface is flattened.
At time $t$ if $F(x,t) < \eta^{th}\left(x,u(x,t)\right)$ the point is stable while if $F(x,t) \geq \eta^{th}\left(x,u(x,t)\right)$ the point is unstable~: it topples, namely
$u(x, t+1) = u(x,t)+1$, and the new threshold $\eta^{th}(x,u+1)$ is drawn.
Avalanches are produced in a quasi-static protocol~: 
when all points are stable $w$ is increased up to a value at which a first instability occurs
and the dynamics unfolds until the line reach a new stable configuration.
We focus on the stationary regime of avalanches, reached after a transient when starting from an arbitrary configuration.
Our simulations are performed for a line of size $L=2^{17}$ for $\alpha=0.5$, 0.75, 1 and 1.5 and many values of the curvature for each $\alpha$. The details of the parameters are given in the Supplemental Material~\citep{SupMat}.
Note that the positive curvature $m^2$ sets the distance from the critical point (at $m=0$) and controls the cutoffs of all power law distributions.
It corresponds to $m^2 \sim 1-R_0$ in the BGW model.

This model was implemented in Ref.~\cite{laurson2010} for $\alpha=1$ and the analysis of the data led to conjecture the following scaling relation for the cluster size exponent $\tau_c$~:
\begin{equation}
\tau_c = 2 \tau - 1 \, . \label{eq: tau_c=2tau-1} 
\end{equation}
The justification of this relation was based on two assumptions~:
(i) During the avalanche spreading the evolution of the number of clusters is interpreted as a 
discrete random walk with the time replaced by the number of topplings.
At each time step, the number of clusters can increase by $1$, remain constant, or decrease by $1$.
This implies that the average number of clusters at the end of avalanches of size $S$ (equal to the total number of topplings) scales as
$\langle N_c(S) \rangle \sim S^{\gamma_S}$.
The assumption that the process is Markovian leads to 
$\gamma_S=1/2$. 
(ii) The clusters belonging to an avalanche of size $S$ have a typical size 
$S_c \sim S / \langle N_c(S) \rangle \sim S^{1 - \gamma_S}$
and applying $P(S_c) dS_c = P(S)dS$ one finds~\cite{footnote_1}~:
\begin{equation}
\tau_c = \frac{\tau - \gamma_S}{1 - \gamma_S} \, , \label{eq: Zapperi conj. generalized}
\end{equation}
which yields \eqref{eq: tau_c=2tau-1} for $\gamma_S=1/2$.
In the case $\alpha=1$, the value of $\gamma_S$ was found  numerically to be compatible with $1/2$. 
In figure~\ref{Fig: 2} we extend the study to $1/2 \leq \alpha < 2$.
We find that in general 
$\gamma_S \neq 1/2$ and that it 
decreases continuously with $\alpha$.
Inserting our measured values of $\gamma_S$ in \eqref{eq: Zapperi conj. generalized} yields predictions for $\tau_c$ that differ from the observed values (e.g. for $\alpha=0.75$ we measured $\gamma_S=0.73$ and $\tau_c=1.72$
while~\eqref{eq: Zapperi conj. generalized} yields $\tau_c=2.33$).
Equation~\eqref{eq: Zapperi conj. generalized} is wrong as the assumption (ii) is incorrect~:
clusters belonging to avalanches of size $S$ do not have the same typical size but are broadly
distributed according to a power law distribution
$P(S_c|S) \sim S_c^{-\tau_c}$ for $S_c$ up to $S$. This implies
$\langle N_c(S) \rangle \sim S^{\tau_c-1}$ and thus $\tau_c = \gamma_S+1$ instead of~\eqref{eq: Zapperi conj. generalized} (see~\cite{SupMat}).
The exponent $\gamma_S$, which varies continuously with $\alpha$, is not independent but is related to $\tau$ as we demonstrate by studying the distribution of the cluster number $N_c$.

\begin{figure}[t]
\includegraphics[width=1\linewidth]{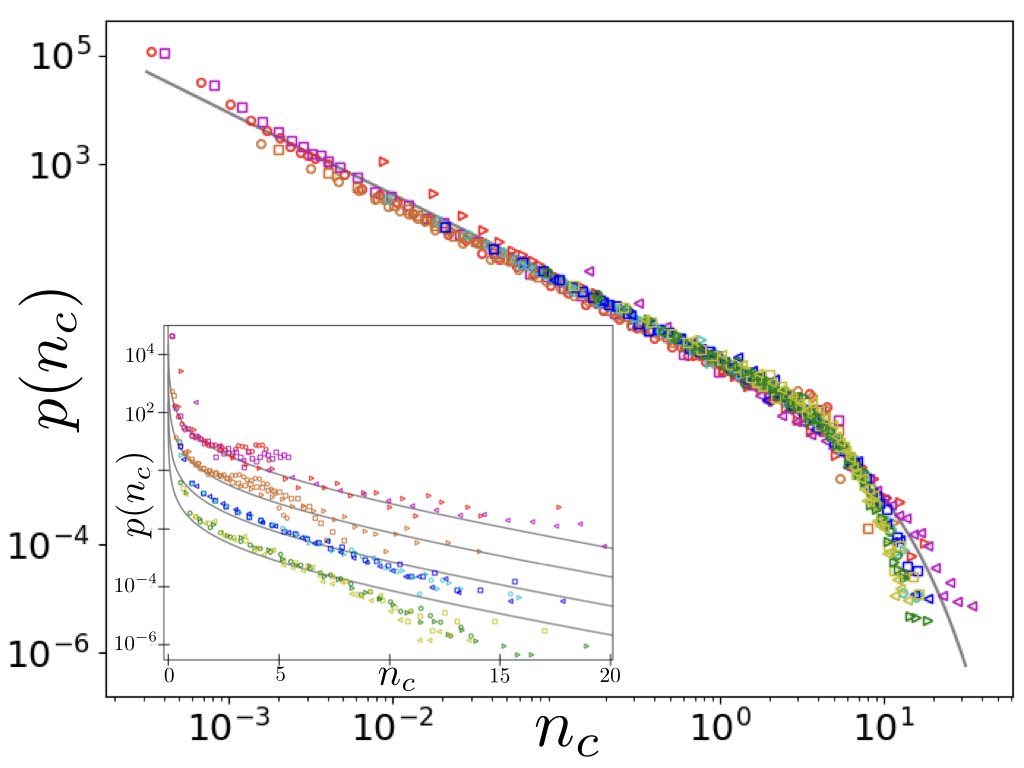}
\caption{\textbf{Main panel :} $p(n_c)$ (as defined in equation \eqref{eq: def p(n_c) 1}) for $\alpha=0.5$ (magenta), 0.75 (brown), 1 (blue) and 1.5 (yellow). \textbf{Inset~:} Same data in log-lin scale and with a vertical shift of the data to test the exponential decay. 
 The thin gray lines correspond to the BGW function $p_{\text{BGW}}$ \eqref{eq: Galton-Watson nc}
 \label{Fig: 3}}
\end{figure}

In figure~\ref{Fig: 3} we study the statistics of the cluster number
for different values of $\alpha$ and $m^2$.
In order to produce a collapse for different values of $m^2$ we adopt the procedure described in \cite{rosso2009}~: we introduce the variable
$n_c = 2N_c \times \langle N_c \rangle / \langle N_c^2 \rangle $
and compute the function $p(n_c)$ defined as~:
\begin{equation}\label{eq: def p(n_c) 1}
p(n_c)dn_c = \frac{\langle N_c^2 \rangle}{2\langle N_c \rangle^2} P\left(N_c\right) dN_c \, .
\end{equation}
For small values of the curvature, $p(n_c)$ is fully universal and its analytical form is known for the BGW process and reads~\cite{ledoussal2009, ledoussal2009b}~:
\begin{equation}\label{eq: Galton-Watson nc}
p_{\text{BGW}}(n_c) = \frac{n_c^{-3/2}}{2\sqrt{\pi}} \exp(-n_c/4) \, .
\end{equation}
Remarkably the data seem to collapse on the same universal curve for all $\alpha$ and 
cannot be clearly distinguished from the BGW function of equation 
\eqref{eq: Galton-Watson nc}.
In the case $\alpha=0.5$ one expects that the number of clusters is proportional to the avalanche size (up to logarithmic corrections). Thus it is not too surprising to recover a BGW shape in this case, since, in mean-field, this is the distribution of the avalanche size.
It is however much less clear why this behavior should hold for $\alpha > 0.5$. Numerically we cannot
exclude that $p(n_c)$ is BGW (in the limit $m \to 0$, $L\to \infty$) also for $\alpha=0.75$, $\alpha=1$ and $\alpha=1.5$.
However, for what follows we need only to assume that $P(N_c)\sim N_c^{-3/2}$ and 
we do not mind about the exact decay at large scale.
To conclude the argument it is sufficient to remark that $P(N_c) dN_c = P(S) dS$ (see~\cite{SupMat} where we show that $P(N_c|S)$ is peaked around $\langle N_c \rangle_S$) which yields 
$\gamma_S=2(\tau-1)$. Combined with the relation $\tau_c=\gamma_S+1$ obtained
above, we arrive at the scaling relation~\eqref{eq: tau_c=2tau-1} 
which we thus find to be valid for all $\alpha$, by a completely different mechanism from \cite{laurson2010}.
Note that we can also infer that the cutoff of $P(N_c)$ scales as $N_m \sim S_m^{\tau_c-1}$.

The same reasoning can be repeated to find the relation between the extension $\ell$ of an avalanche and the extension $\ell_c$ of the clusters.
Using the definition for $\gamma_{\ell}$, $\langle N_c \rangle_{\ell} \sim \ell^{\gamma_{\ell}}$,
we first derive the relation $\gamma_{\ell} = \kappa_c-1$.
Then assuming that the distribution of $N_c$ originates from a BGW process we derive~:
\begin{equation}
\kappa_c = 2\kappa -1 \, , \label{eq: kappa_c = 2 kappa - 1}
\end{equation}
by using $P(N_c) dN_c = P(\ell) d\ell$.
Note that from $\gamma_S = \tau_c - 1$, $\gamma_{\ell} = \kappa_c-1$ and $S\sim \ell^{d+\zeta}$ we can also derive $\kappa_c-1 = (\tau_c-1)(d+\zeta)$ which corresponds to the self-affine property of the clusters
$S_c \sim \ell_c^{d+\zeta}$ (see~\cite{SupMat}).

\begin{figure}[t]
	\includegraphics[width=1\linewidth]{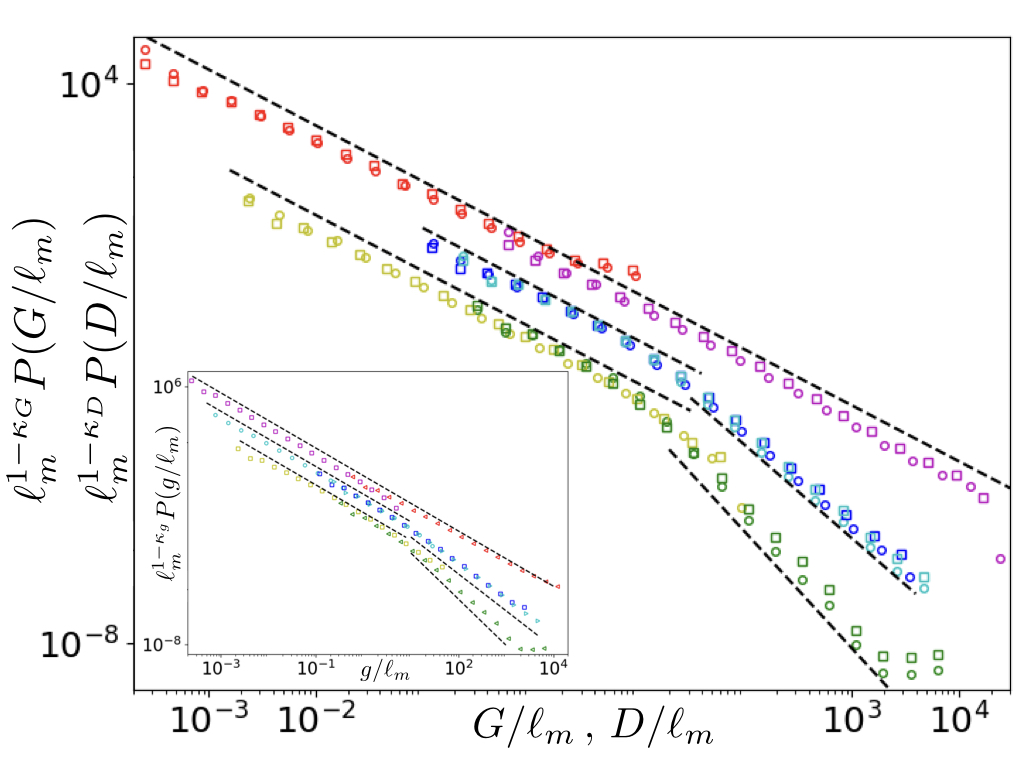}
	\caption{\textbf{Main panel :} Diameter (circles) and total gap length (squares) distributions for $\alpha=0.5$, 1 and 1.5. $P(D/\ell_m)$ is the probability distribution of the variable $D/\ell_m$ with $\ell_m = m^{-2/\alpha}$. Dashed lines are fit with exponents $\kappa_D=1.2$ at small scales and $d+\alpha$ at large scale for $\alpha=1$ and 1.5.
	Note that the distribution of $D$ and $G$ are indistinguishable.
	\textbf{Inset~:} Gap distributions.
	The dashed lines are guide to the eyes of exponents $\kappa_g=3/2$ at small scale and $d+\alpha$ at large scale. 
	See~\cite{SupMat} for details of the parameters.
	\label{Fig: 4}}
\end{figure}

The scaling relations that we have derived allow to express all the exponents as functions of $\zeta$ and $\alpha$. We have collected all these expressions in table~\ref{tab: exponents} and compared the predicted values of the exponents with the ones directly measured in our numerical simulations. 
We found good agreements for all values of $\alpha$.

Note that at variance with SR elasticity the diameter $D$ of the region that is perturbed by the presence of an avalanche does not coincide with the avalanche linear size $\ell$. 
The diameter is in general much larger than $\ell$ and displays novel statistical properties that we should characterize.

In figure~\ref{Fig: 4} we show the diameter distribution for $\alpha=0.5$, 1 and 1.5.
For $\alpha > 0.5$ a crossover $ D_{\text{cross}} \sim \ell_m $ separates two power law decays~:
\begin{align}
P(D) &\sim \left\lbrace 
\begin{array}{l}
D^{-\kappa_D} \, \text{ for } \, D < D_{\text{cross}} \, , \\
D^{-(d+\alpha)} \, \text{ for } \, D > D_{\text{cross}} \, .
\end{array}\right. \label{eq: P(D)}
\end{align}
Remarkably the exponent $\kappa_D$ is $\alpha$ independent with a numerical value close to $1.2$ (also for $\alpha=0.5$). 
Hence the statistics of the diameter is very different from the one of the linear size $\ell$.
Since $D = \ell + G$, where $G = \sum g$ is the total gap, we need to investigate the statistics of the gaps $g$.
Their distributions, shown in the inset of figure~\ref{Fig: 4}, present a crossover $g_{\text{cross}} \simeq 10 \ell_m$ between two power law decays~:
\begin{align}
P(g) &\sim \left\lbrace 
\begin{array}{l}
g^{-\kappa_g} \, \text{ for } \, g < g_{\text{cross}} \, , \\
g^{-(d+\alpha)} \, \text{ for } \, g > g_{\text{cross}} \, ,
\end{array}\right. \label{eq: P(g)}
\end{align}
with $\kappa_g=3/2$ independently from $\alpha$.
Since the decay of $P(g)$ is much slower than the one of $P(\ell_c)$ (which decays faster than any power law at large scale, see~\cite{SupMat}),
we expect $G \gg \ell$ which implies that $D\simeq G$. 
This latter relation is confirmed by figure~\ref{Fig: 4} where the distribution of $D$ and $G$ are plotted together and are indistinguishable.

The scaling form~\eqref{eq: P(D)} can now be derived from the gap distribution~\eqref{eq: P(g)}. Let us first consider avalanches where all gaps are below $g_{\text{cross}}$. In this case we can write~:
\begin{equation}
\langle N_c \rangle_G = 1 + \frac{G}{\langle g \rangle_G } \sim G^{\kappa_g-1} \simeq D^{\kappa_g-1} \, , \label{eq: relation Nc vs G}
\end{equation}
where we used $\langle g \rangle_G \sim G^{2-\kappa_g}$ since 
$P(g|G)\sim g^{-\kappa_g}$ for $g$ up to $G$ (see~\cite{SupMat}).
Again using $P(N_c)\sim N_c^{-3/2}$ and setting $P(N_c)dN_c = P(D)dD$ we find~:
\begin{equation}
\kappa_D = (\kappa_g + 1)/2 \simeq 1.25 \, , \label{eq: conj. kappa_D}
\end{equation}
which is close from the exponent we measure $\kappa_D \simeq 1.2$.
A very large value of $D$ corresponds to the presence of one (or a very few) gap $g > g_{\text{cross}}$. This gap has a probability $\propto g^{-(d+\alpha)}$ and dominates $D$ 
which hence also has a probability $\propto D^{-(d+\alpha)}$.

\smallskip
The main result of this paper is to show that, even in presence of LR interactions, the statistical properties of depinning avalanches can always be expressed as functions of the elasticity range parameter $\alpha$ and the roughness exponent $\zeta$. 
Our conclusions are based on the numerical observation that the number of clusters behave as the number of offsprings of a BGW model for all $\alpha$ between $1/2$ and $2$. 
We are not able to demonstrate this conjecture but our numerics shows that it is a very good approximation.  
All relations between the exponents of clusters and global avalanches
are based on this conjecture so it would be of great interest to show it analytically.
It would be also interesting to characterize the statistical properties of the clusters of plastic avalanches~\cite{baret2002, lin2014, salerno2012, nicolas2018}. Indeed the Eshelby kernel, which is the relevant interaction for the yielding transition,
is long-range (with $\alpha=0$) but, in contrast with depinning, non-monotonous.

\begin{acknowledgments}
{\it Acknowledgments:}
We thank L. Ponson for enlightening discussions.
PLD acknowledges support from ANR under the grant ANR- 17-CE30-0027-01 RaMaTraF
\end{acknowledgments}

\onecolumngrid
\FloatBarrier
\begin{adjustwidth}{-2cm}{-2cm}
\begin{center}
\begin{table*}
	\begin{tabular}{c c c c c c c}
		 & & & \multicolumn{4}{c}{measured\, /\, prediction}  \\
		Exponent & Expression & Relation & $\alpha=0.5$ & $\alpha=0.75$ & $\alpha=1$ & $\alpha=1.5$ \\	
		\hline
		$\zeta$ & $S(q)\sim q^{-(d+2\zeta)}$ &  & 0 (MF) & 0.18(1) & 0.39~\cite{rosso2002} & 0.75(2) \\ 
		$\gamma_S$ & $\langle N_c \rangle_S \sim S^{\gamma_S}$ & $\gamma_S = 2-\frac{2\alpha}{d+\zeta}$ &
		\, 0.89(2)\, /\, 1 \,\; & \, 0.73(1)\, /\, 0.73(1) \,\; & \, 0.52(2)\, /\, 0.56 \; & 0.29(1)\, /\, 0.28(2) \\ 
		$\gamma_{\ell}$ & $\langle N_c \rangle_{\ell} \sim S^{\gamma_{\ell}}$ & 
		$\gamma_{\ell}=2(d+\zeta - \alpha)$ & 
		\, 0.93(2)\, /\, 1\, ; & \, 0.80(2)\, /\, 0.86(2) \,\; & \, 0.70(2)\, /\, 0.78 \;  & 0.50(2)\, /\, 0.54(4) \\
		$\tau$ & $P(S)\sim S^{-\tau}$ & $\tau = 2-\frac{\alpha}{d+\zeta}$ & 1.50(1)\, /\, $\frac{3}{2}$ & 
		1.36(2)\, /\, 1.36(1) &	1.26(2)\, /\, 1.28 & 1.14(2)\, /\, 1.14(1) \\ 
		$\tau_c$ & $P(S_c)\sim S^{-\tau_c}$ & $\tau_c = 3-\frac{2\alpha}{d+\zeta}$ & 
		2.00(9)\, /\, 2\; & 1.72(4)\, /\, 1.73(1) & 1.56(2)\, /\, 1.56 & 1.28(2)\, /\, 1.28(2) \\ 
		$\kappa$ & $P(\ell)\sim S^{-\kappa}$ & $\kappa = 1+d+\zeta-\alpha$ & 1.50(2)\, /\, $\frac{3}{2}$ & 
		1.38(5)\, /\, 1.43(1) & 1.33(5)\, /\, 1.39 & 1.20(3)\, /\, 1.25(2) \\ 
		$\kappa_c$ & $P(\ell_c)\sim S^{-\kappa_c}$ & \; $\kappa_c = 1+2(d+\zeta-\alpha)$\; & 
		2.05(5)\, /\, 2 & 1.90(5)\, /\, 1.86(2) & 1.80(2)\, /\, 1.78 & 1.45(3)\, /\, 1.50(4) \\ \hline
	\end{tabular}
	\caption{Table of exponents. The measured values correspond to best fits of our data. The predictions correspond to the scaling relations indicated in the third column, for $d=1$, using the values of $\zeta$ listed in the first line.
	 Note that we use the mean-field (MF) value $\zeta=0$ for $\alpha=0.5$ and a previous, more precise, numerical determination $\zeta=0.39$ for $\alpha=1$. 
	 \label{tab: exponents}}
\end{table*}
\end{center}
\end{adjustwidth}
\FloatBarrier
\twocolumngrid

\bibliography{BiblioClusters}

\begin{thebibliography}{46}
\bibitem[{foo()}]{footnote_1}
\bibinfo{note}{For convenience, here and below we use the same letter $P$ to
  denote the probability densities of all observables.}
\end{thebibliography}

\appendix


\begin{widetext}

\bigskip

\bigskip

\begin{large}
\begin{center}

Supplemental Material for {\it Spatial clustering of depinning avalanches in presence of long-range interactions}

\end{center}
\end{large}

In this supplemental material we give the details of the analysis of our extensive numerical simulations.
In appendix A we give the details of the parameters used for the simulations.
In appendix B we detail the measurement of the exponents describing the avalanche statistics. 
Finally, in appendix C we verify the assumptions made in our argument in the main text.

\subsection{A) Details of the parameters}

Figure~\ref{fig: ColorCode} summarizes the values of the parameters used in the simulations as well as the color code used in all figures except figure~\ref{Fig: 4} of the main text and figure~\ref{fig: P(S_c|S) and P(ell_c|ell)} of the supplemental material.

\vspace{1cm}
\begin{figure}[h]
	\centering
	\includegraphics[width=.8\linewidth]{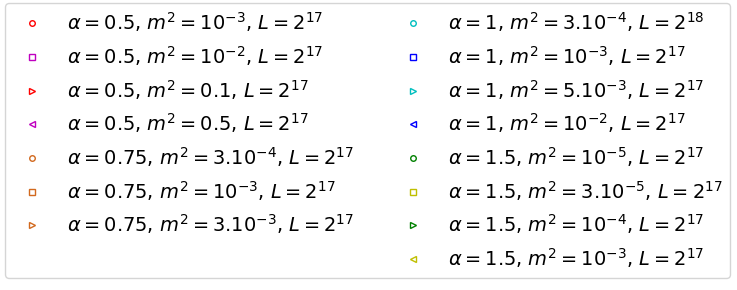}
	\caption{Color code for all figures except figure~\ref{Fig: 4} of the main text and figure~\ref{fig: P(S_c|S) and P(ell_c|ell)} of the supplemental material.
	Red and magenta are used for $\alpha=0.5$, brown for $\alpha=0.75$, cyan and blue for $\alpha=1$ and yellow and green for $\alpha=1.5$.  \label{fig: ColorCode}} 
\end{figure}

The parameters used in figure~\ref{Fig: 4} of the main text are given below while the ones used in figure~\ref{fig: P(S_c|S) and P(ell_c|ell)} of the Supplemental Material are specified in the caption.

\paragraph{\textbf{Main panel of figure~\ref{Fig: 4} $\left(P(D)\text{ and }P(G) \right)$~:}}
$\alpha=0.5$ : $L=2^{17}$ and $m^2=0.01$ (red symbols) and $L=2^{17}$ and $m^2=0.5$ (magenta) ; 
$\alpha=1$ : $L=2^{17}$ and $m^2=0.05$ (blue) and $L=2^{17}$and $m^2=0.1$ (cyan) ;
$\alpha=1.5$ : $L=2^{17}$ and $m^2=3.10^{-5}$ (yellow) and $L=2^{17}$ and $m^2=0.03$ (green).

\paragraph{\textbf{Inset of figure~\ref{Fig: 4} $\left(P(g)\right)$~:}}
$\alpha=0.5$ : $L=2^{17}$ and $m^2=0.01$ (magenta triangles) and $L=2^{17}$ and $m^2=0.5$ (red squares) ; 
$\alpha=1$ : $L=2^{18}$ and $m^2=3.10^{-4}$ (cyan circles), $L=2^{17}$ and $m^2=0.05$ (blue triangles) and $L=2^{17}$and $m^2=0.1$ (cyan squares) ;
$\alpha=1.5$ : $L=2^{17}$ and $m^2=3.10^{-5}$ (yellow squares) and $L=2^{17}$ and $m^2=0.03$ (green triangles).

\subsection{B) Measurement of the exponents}

\begin{figure}[b]
	\centering
	\includegraphics[scale=0.25]{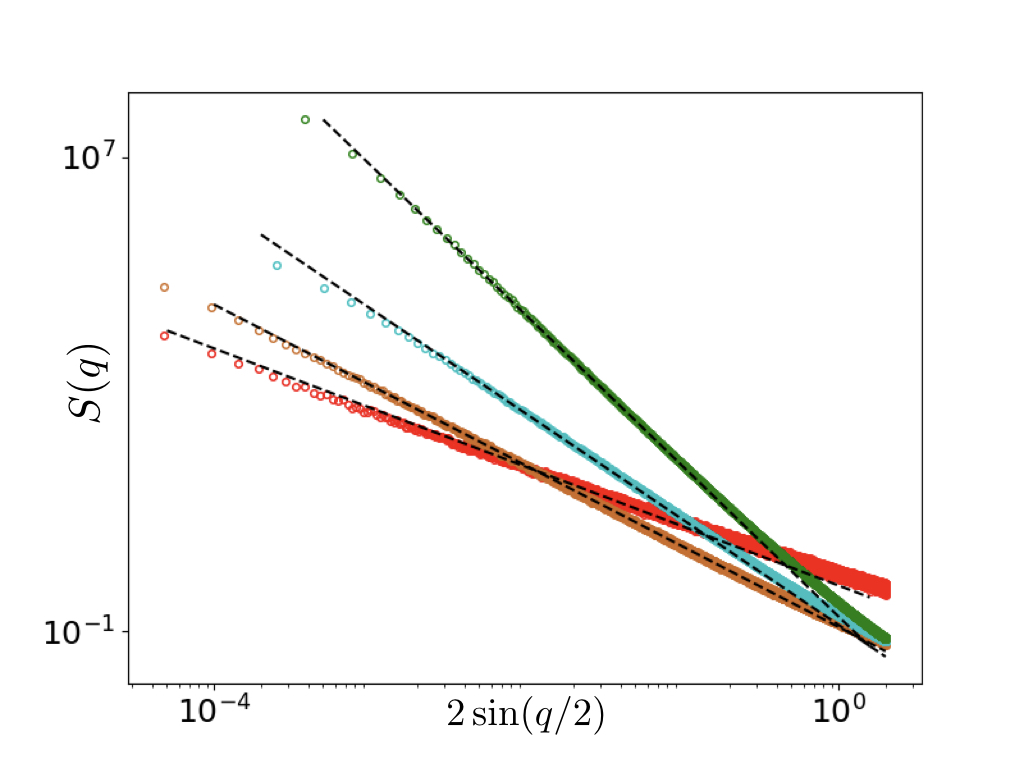}
	\caption{Structure factor. 
	The dashed lines correspond to the fit $q^{-(1+2\zeta)}$ with the values of $\zeta$ given in table~\ref{tab: comparison zeta}.
	The color code is the one given in figure \ref{fig: ColorCode} except that $L=25.10^{3}$ for $\alpha=1$ and $L=2^{14}$ for $\alpha=1.5$.\label{fig: Roughness}} 
\end{figure}

In this section we present the analysis of our extensive numerical simulations in order to measure
directly the exponents of the avalanche statistics.
All the measurements with error margins are summarized in table~\ref{tab: measurement with errors}.

\paragraph{\textbf{Roughness exponent}}
The roughness exponent $\zeta$ is determined from the recorded stable configurations $u(x)$, using the structure factor~:
\begin{align}
S(q) &= \overline{u(q) u(-q)} \sim q^{-d+2\zeta)} \, , \label{eq: S(q)} \\
u(q) &= \int_{-\infty}^{+\infty} \frac{dq}{2\pi} e^{-iqx} u(x) \, , \label{eq: TF u(q)}
\end{align}
with $d=1$ the internal dimension of the interface.
The overline denotes average over disorder. In practice the average is performed over many stable configurations visited by the line during the simulation.
Equation~\eqref{eq: S(q)} is valid for $0<\zeta <1$. Slopes smaller than $q^{-d}$ correspond to $\zeta=0$ with logarithmic corrections. 
We compute $S(q)$ for each $\alpha$, using the smallest values of the curvature $m^2$ used in the simulations.
The results are plotted in figure~\ref{fig: Roughness}.

Our best fit values of $\zeta$ are given in table~\ref{tab: comparison zeta} where they are compared with the one loop and two loop calculations from the Functional RG, which are (presumably divergent) series expansions in 
$\epsilon=2 \alpha-d$ to order $O(\epsilon)$ and $O(\epsilon^2)$ respectively.
Setting simply $d=1$ in these series the one loop result is $\zeta^{1\text{loop}} = \frac{2\alpha-1}{3}$~\cite{ertas1994, ledoussal2002}. The two loop result reads 
$\zeta^{2\text{loop}} = \zeta^{1\text{loop}} + \frac{(2\alpha-1)^2}{3}\frac{\psi(1)+\psi(\alpha)- 2\psi(\alpha/2)}{9\sqrt{2} \gamma}$, from formula 
(4.17) and (3.60) in Ref.~\cite{ledoussal2002} (we noted that the integral in (3.60) can be performed explicitly). Note that these formula can also be seen as an expansion around $\alpha=1/2$ for fixed $d=1$.
For $\alpha=1$ there exists a previous more precise numerical determination $\zeta=0.388(2)$~\cite{rosso2002}.
It is this value that we used for computing the numerical values of the predictions in the table of the main text.

\begin{center}
\begin{table}[h]
	\begin{tabular}{c c c c c c}
		  &  $\alpha=0.5$ & $\alpha=0.75$ & $\alpha=1$ & $\alpha=1.5$ \\ 		
		\hline
		$\zeta$ from Fig.~\ref{fig: Roughness} & $\sim 0$ & $0.18(1)$ & $0.37(2)$ & 
		$0.77(2)$ \\
		$\zeta^{1\text{loop}}$ & 0 & $1/6$ & $1/3$ & $2/3$ \\ 
		$\zeta^{2\text{loop}}$ & 0 & 0.21 & 0.46  & 0.96\\
		\hline		
	\end{tabular}
	\caption{Comparison of our measurements of the roughness exponent with the one loop and two loop predictions. As noted in \cite{ledoussal2002} for $\alpha=1$ and 2 (see table 2 there), the numerical value is intermediate between the 1 loop and 2 loop estimates
	\label{tab: comparison zeta}}
\end{table}
\end{center}

\paragraph{\textbf{Avalanche exponents}}
We now focus on the exponents of the global avalanche. We expect the avalanche size $S$ and linear size $\ell$ to obey the following distributions~:
\begin{align}
P(S) &= S^{-\tau} f_S \left( S/S_m \right) \, , \label{eq: P(S)} \\
P(\ell) &= \ell^{-\kappa} f_{\ell} \left( \ell/\ell_m \right) \, , \label{eq: P(ell)}
\end{align}
where $f_S(y)$ and $f_{\ell}(y)$ are some scaling functions that are constant for $y \ll 1$ and present a fast decay for $y\gg 1$.
The cutoff scales $S_m$ and $\ell_m$ are controlled by the harmonic confinement curvature $m^2$.
It induces a characteristic length $\ell_{m} = m^{-2/\alpha}$ beyond which the line is flattened.
This length is a natural cutoff for the avalanche linear size distribution. Thanks to the self-affinity property of the avalanches it also sets the cutoff for the total avalanche size
$ S_{m} = \ell_{m}^{d+\zeta} = m^{-2(d+\zeta)/\alpha} $.

The avalanche global and linear size distributions are plotted in figure~\ref{fig: P(S) and P(ell)}. The dashed lines correspond to fits of the exponent $\tau$ and $\kappa$ listed in table~\ref{tab: measurement with errors}. The data have been rescaled using the cutoff scales $S_m$ and $\ell_m$. 
The collapse of the data for each $\alpha$ is a confirmation of the scaling of the cutoffs as well as of the values of the exponents.

\begin{figure}
    \begin{adjustwidth}{-1cm}{-1cm}
	\centering
	\includegraphics[scale=0.25]{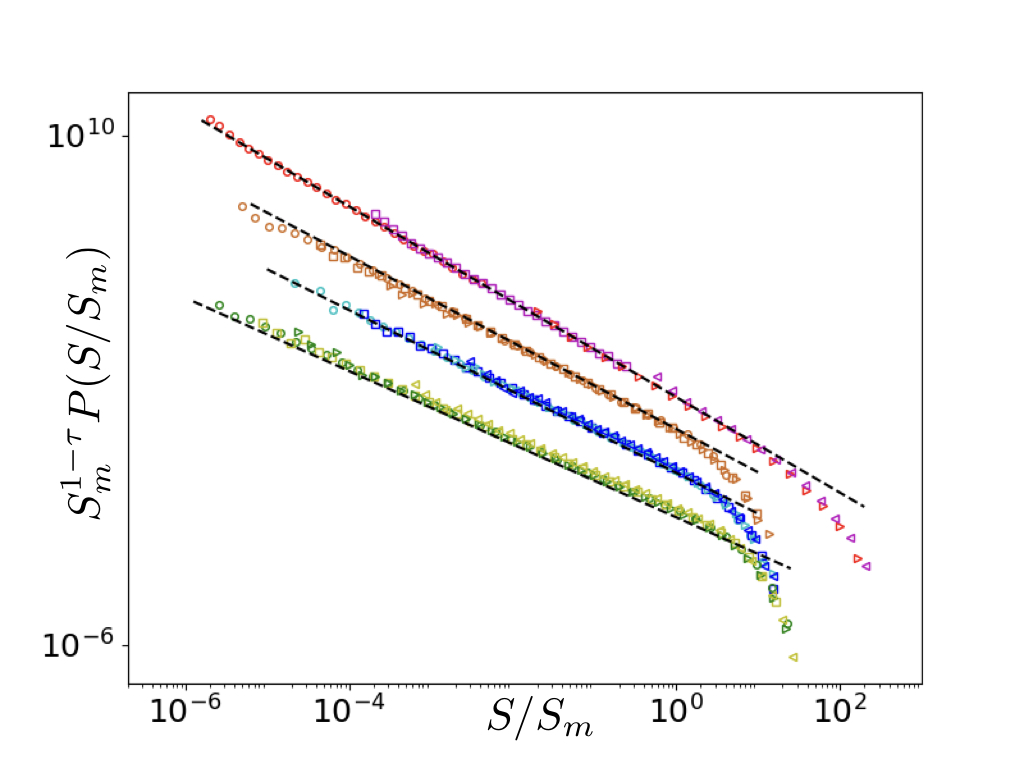}\includegraphics[scale=0.25]{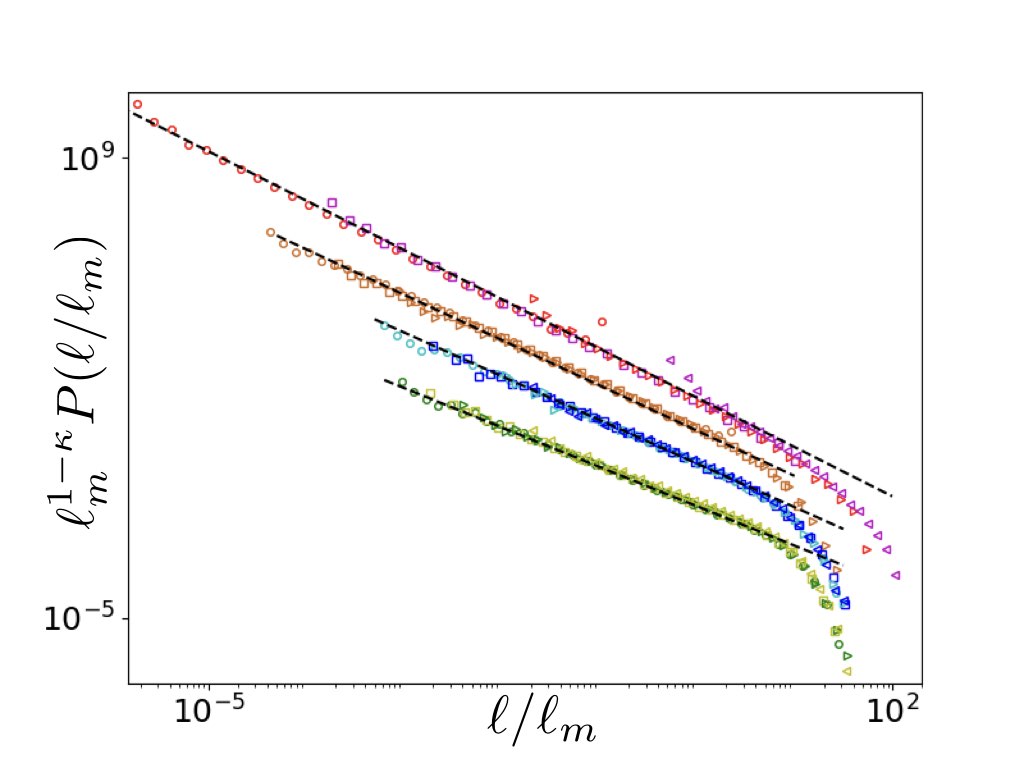}
	\caption{\textbf{Left:} Collapse of the avalanche size distribution. 
	$P(S/S_m)$ is the probability distribution of the variable $S/S_m$ with $S_m = \ell_m^{d+\zeta} = m^{-2(d+\zeta)/\alpha}$. The dashed line are fits of the exponent $\tau$ with the values listed in table~\ref{tab: measurement with errors}.
	\textbf{Right :} Collapse of the avalanche linear size distribution. $P(\ell/\ell_m)$ is the probability distribution of the variable $\ell/\ell_m$ with $\ell_m = m^{-2/\alpha}$. The dashed line are fits of the exponent $\kappa$ with the values listed in table~\ref{tab: measurement with errors}
	Data have been shifted vertically to enhance visibility.
	\label{fig: P(S) and P(ell)}} 
	\end{adjustwidth}
\end{figure}

\paragraph{\textbf{Definition of the cluster distribution}}
In this paper we define the distribution of cluster size $S_c$ as follows. We take 
$M$ avalanches, which have different cluster number $N_c^{i}$, and let $S_c^{i, j}$ be the $j$th cluster of avalanche $i$. Then taking the limit $M \to \infty$ we define~:
\begin{equation}\label{eq: def P(S_c)} 
P(S_c) = \lim_{M \to \infty}\frac{1}{\sum_{i=1}^{M} N_c^{i}} \sum_{i=1}^{M} \sum_{j=1}^{N_c^{i}} \delta(S_c^{i,j} - S_c) 
\end{equation}
This prescription amounts to pick one cluster at random among all clusters of all avalanches.
The definition of the conditional probability $P(S_c|S)$ is the same where one selects only avalanches
of size $S$. 
From our definition we can derive the equality 
$\langle S_c \rangle_S  = S/\langle N_c \rangle_S$
used in the main text.
Indeed considering $M$ avalanches of equal size $S$ the average of the cluster sizes over these avalanches is~:
\begin{equation}\label{eq: equality S_c average}
\langle S_c \rangle_S = \frac{1}{\sum_{i=1}^{M} N_c^{i}} \sum_{i=1}^{M} \sum_{j=1}^{N_c^{i}}S_c^{i,j}
 = \frac{M\times S}{\sum_{i=1}^{M} N_c^{i}} = \frac{S}{\langle N_c \rangle_S}
\end{equation}

The distributions  of cluster extensions $P(\ell_c)$ and of gaps $P(g)$ are defined similarly as in equation~\eqref{eq: def P(S_c)}.
Note that for the gaps $N_c^{i}$ must be replaced by $N_c^{i}-1$ because there are $N_c-1$ gaps for $N_c$ clusters.
In the main text we used the equalities
\begin{align}
\langle \ell_c \rangle_{\ell} = \frac{\ell}{\langle N_c \rangle_{\ell}} \, , \label{eq: equality ell_c} \\
\langle g \rangle_G  = \frac{G}{\langle N_c \rangle_G -1} \, , \label{eq: equality g}
\end{align}
that can be derived in the same manner as in equation~\eqref{eq: equality S_c average}.

\paragraph{\textbf{Cluster exponents}}
We expect the cluster size $S_c$ and cluster extension $\ell_c$ to obey the following distributions~:
\begin{align}
P(S_c) &= S^{-\tau_c} f_{S_c} \left( S_c/S_m \right) \, , \label{eq: P(S_c)} \\
P(\ell_c) &= \ell^{-\kappa_c} f_{\ell_c} \left( \ell_c/\ell_m \right) \, , \label{eq: P(ell_c)}
\end{align}
with scaling functions $f_{S_c}(y)$ and $f_{\ell_c}(y)$ that are constant for $y \ll 1$ and decay fast for $y\gg 1$.
Note that we use the same cutoff scales $S_m$ and $\ell_m$ as for the global avalanches.

The cluster size and extension distributions are plotted in figure~\ref{fig: P(S_c) and P(ell_c)}. The dashed lines correspond to fits of the exponent $\tau_c$ and $\kappa_c$ listed in table~\ref{tab: measurement with errors}. The data have been rescaled using the cutoff scales $S_m$ and $\ell_m$. 
The collapse of the data for each $\alpha$ is a confirmation of the scaling of the cutoffs as well as of the values of the exponents.

\begin{figure}
    \begin{adjustwidth}{-1cm}{-1cm}
	\centering
	\includegraphics[scale=0.25]{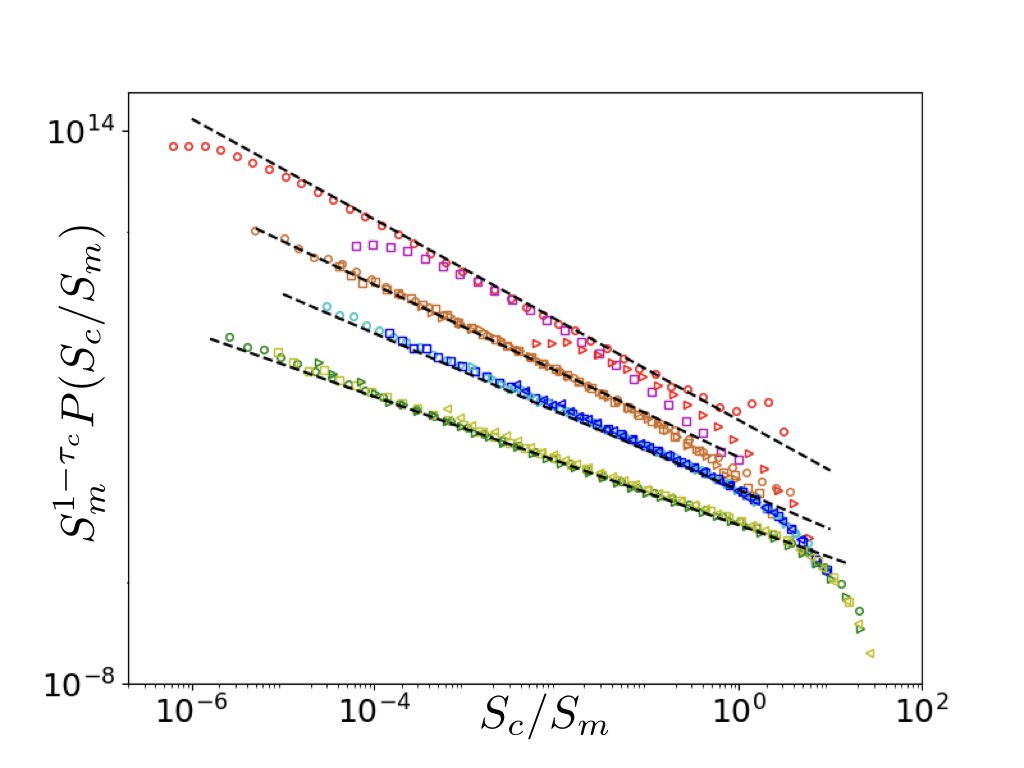}\includegraphics[scale=0.25]{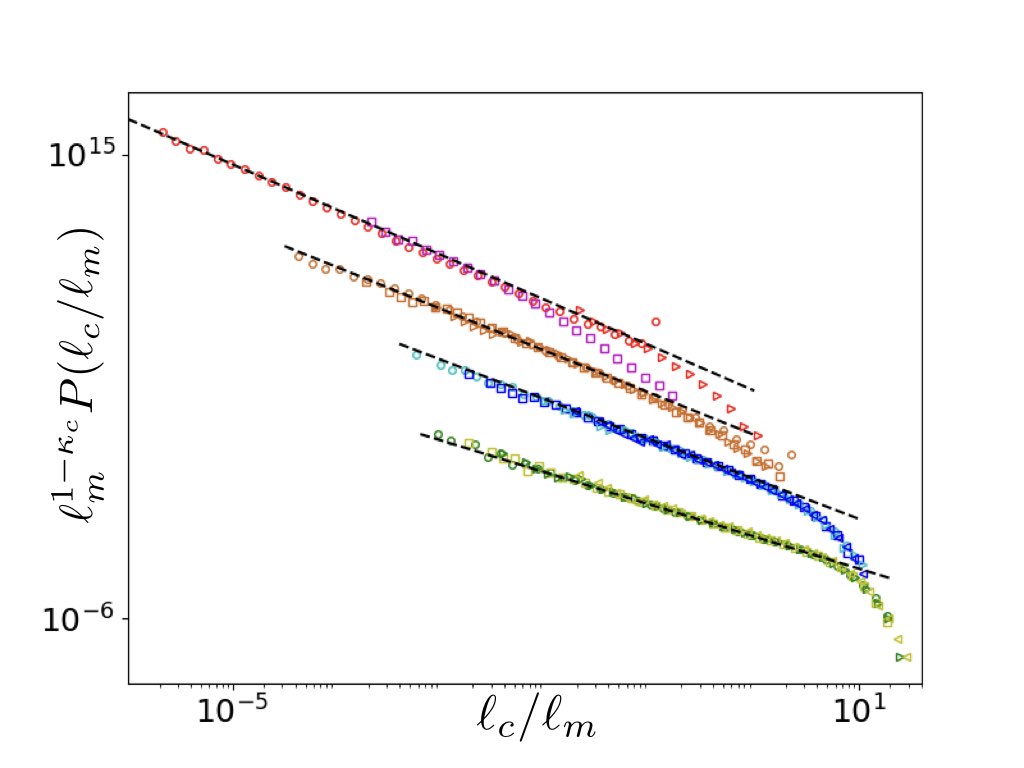}
	\caption{\textbf{Left:} Collapse of the cluster size distribution. 
	$P(S_c/S_m)$ is the probability distribution of the variable $S_c/S_m$ with $S_m = \ell_m^{d+\zeta} = m^{-2(d+\zeta)/\alpha}$. The dashed line are fits of the exponent $\tau_c$ with the values listed in table~\ref{tab: measurement with errors}.
	\textbf{Right :} Collapse of the cluster extension distribution. $P(\ell_c/\ell_m)$ is the probability distribution of the variable $\ell_c/\ell_m$ with $\ell_m = m^{-2/\alpha}$. The dashed line are fits of the exponent $\kappa_c$ with the values listed in table~\ref{tab: measurement with errors}. 
	Data have been shifted vertically to enhance visibility.
	\label{fig: P(S_c) and P(ell_c)}} 
	\end{adjustwidth}
\end{figure}

\begin{center}
\begin{table}
	\begin{tabular}{c c  c c c c}
		Exponent & Expression & $\alpha=0.5$ & $\alpha=0.75$ & $\alpha=1$ & $\alpha=1.5$ \\ 		
		\hline
		$\zeta$ & $S(q)\sim q^{-(d+2\zeta)}$  & $\sim 0$ & $0.18(1)$ & $0.37(2)$ & 
		$0.77(2)$ \\
		$\gamma_S$ & $\langle N_c \rangle_S\sim S^{\gamma_S}$ &
		0.89(1) & 0.73(1) & 0.52(1) & 0.29(1) \\ 
		$\gamma_{\ell}$ & $\langle N_c \rangle_{\ell} \sim S^{\gamma_{\ell}}$ &
		0.93(2) & 0.80(2) & 0.70(2) & 0.50(2) \\
		$\tau$ & $P(S)\sim S^{-\tau}$ & 1.50(1) & 1.36(2) &
		1.26(2) & 1.14(2) \\ 
		$\tau_c$ & $P(S_c)\sim S^{-\tau_c}$ & 2.00(10) &
		1.72(4) & 1.56(2) & 1.28(2) \\ 
		$\kappa$ & $P(\ell)\sim S^{-\kappa}$ & 1.50(2) & 
		1.38(5) & 1.33(4) & 1.20(3) \\ 
		$\kappa_c$ & $P(\ell_c)\sim S^{-\kappa_c}$ &  2.05(5) &
		1.90(5) & 1.80(2) & 1.45(3) \\ 
		$\kappa_D$ & $P(D)\sim S^{-\kappa_D}$ &
		1.20(7) & 1.15(5) & 1.20(10) & 1.21(6) \\ 
		$\tau_{N_c}$ &  $P(N_c) \sim N_c^{-\tau_{N_c}}$ & 1.50(2) & 1.50(3) & 1.50(6) & 1.50(8) \\
		$\kappa_g$ & $P(g)\sim g^{-\kappa_g}$ & 1.50(5) & 1.50(3) & 1.50(5) & 1.50(6) \\ 	
		\hline		
	\end{tabular}
	\caption{Summary of measurements with error margins.
	\label{tab: measurement with errors}}
\end{table}
\end{center}

\subsection{C) Verification of the assumptions}

\begin{figure}
    \begin{adjustwidth}{-1cm}{-1cm}
	\centering
	\includegraphics[scale=0.25]{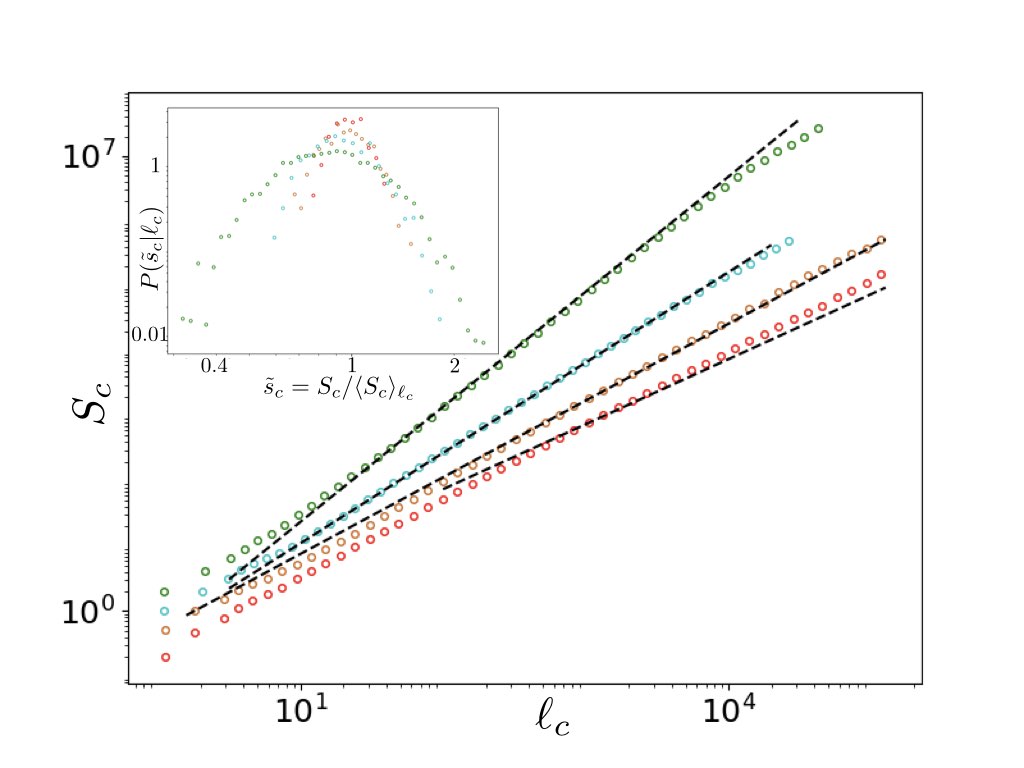}\includegraphics[scale=0.25]{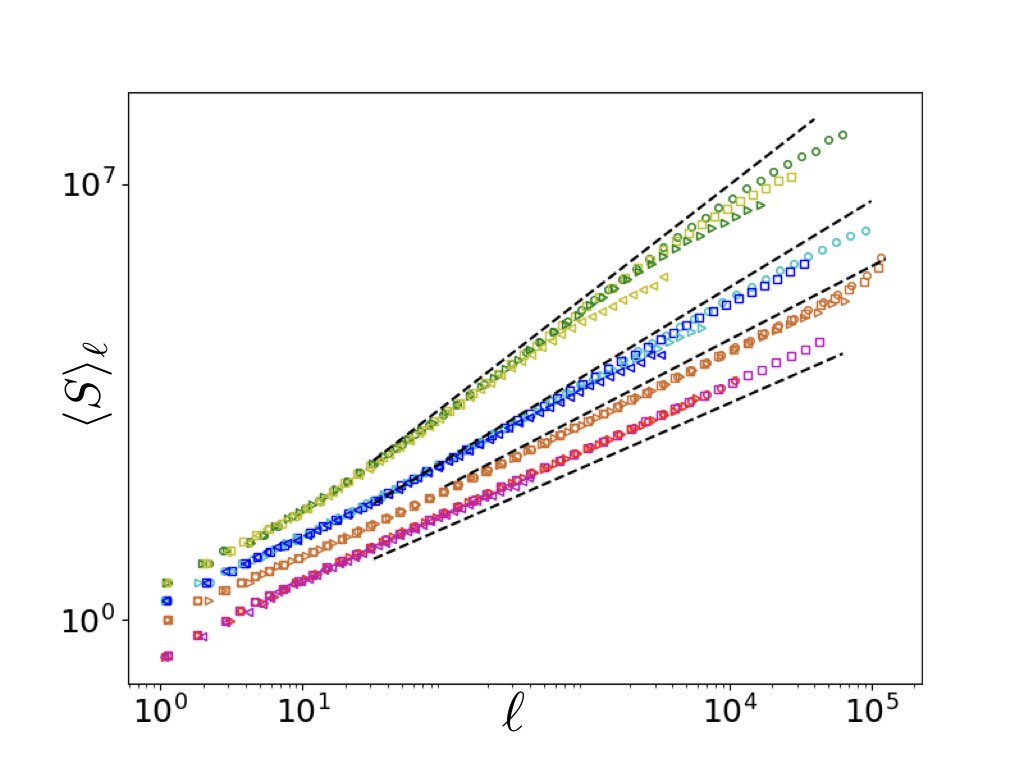}
	\caption{\textbf{Left :} Averaged cluster size $\langle S_c \rangle_{\ell_c}$ versus cluster extension $\ell_c$.
	 Dashed lines correspond to fits $\langle S_c \rangle_{\ell_c} \sim \ell_c^{1+\zeta}$ with the values of $\zeta$ listed in table~\ref{tab: measurement with errors}. \textbf{Inset :} Conditional probability $P(S_c|\ell_c)$ computed in a small interval around $\ell_c = 1300$. The probability is well peaked. This justifies the derivation of equation \eqref{eq: relation kappa_c tau_c zeta}.
	\textbf{Right :} $\langle S \rangle_{\ell}$ versus $\ell$. The dashed lines correspond to the self-affinity scaling $\langle S \rangle_{\ell} \sim \ell^{d+\zeta}$. Data have been shifted vertically to enhance visibility.
	\label{fig: S_c vs ell_c and S vs ell}} 
	\end{adjustwidth}
\end{figure}

In this section we verify the assumptions made in the argument of the main text.
\vspace{.5cm}

\paragraph{\textbf{Self-affinity relations}}
A basic ingredient of the depinning theory is the self-affinity of the interface. 
As a consequence the avalanches display the scaling 
\begin{equation}
S \sim \ell^{d+\zeta} \label{eq: S sim ell d+zeta}
\end{equation}
which is well-verified for short-range avalanches. 
In figure~\ref{fig: S_c vs ell_c and S vs ell} we test that this scaling also holds for long-range avalanches, both for the global avalanches and the clusters, $S_c \sim \ell_c^{d+\zeta}$.
The latter relation is also a consequence of the scaling analysis presented in the main text and yields a scaling relation for the cluster exponents~:
\begin{equation}\label{eq: relation kappa_c tau_c zeta}
\kappa_c -1 = (\tau_c-1)(d+\zeta) \, .
\end{equation}

\begin{figure}
    \begin{adjustwidth}{-1cm}{-1cm}
	\centering
	\includegraphics[scale=0.16]{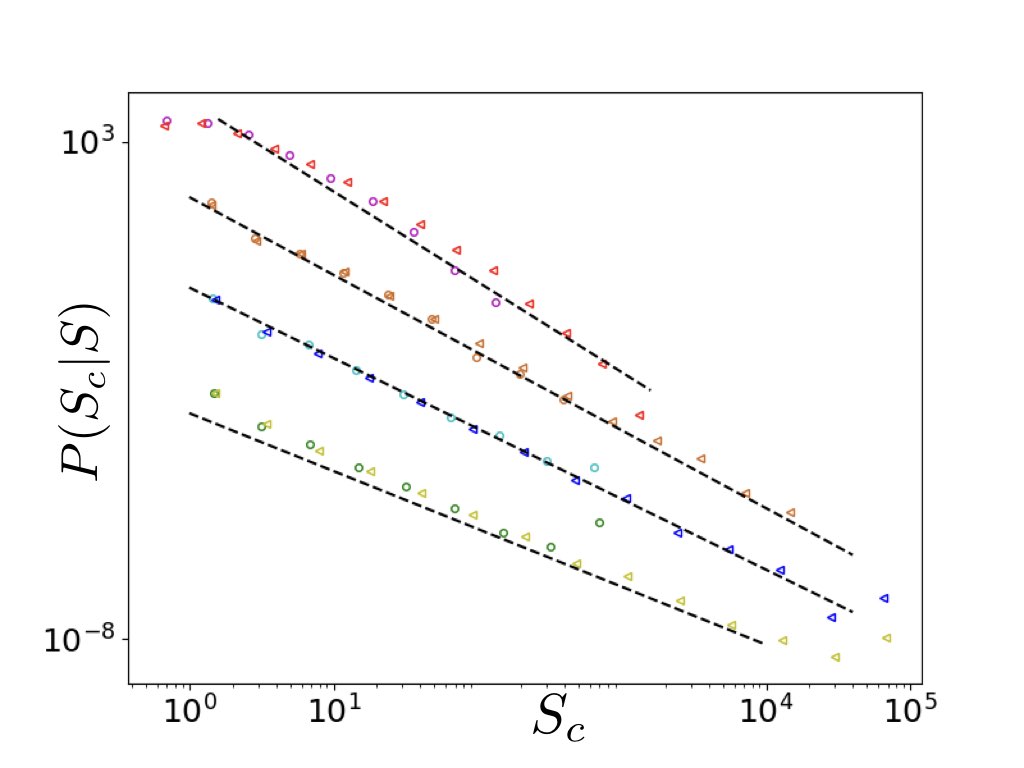}\includegraphics[scale=0.16]{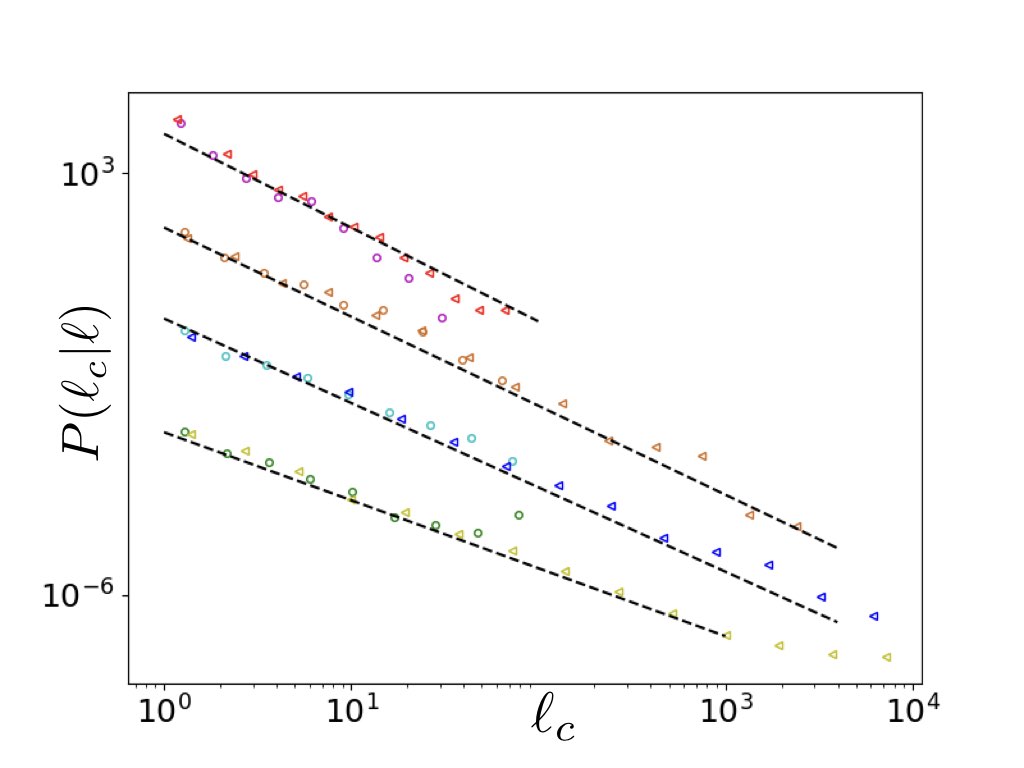}\includegraphics[scale=0.16]{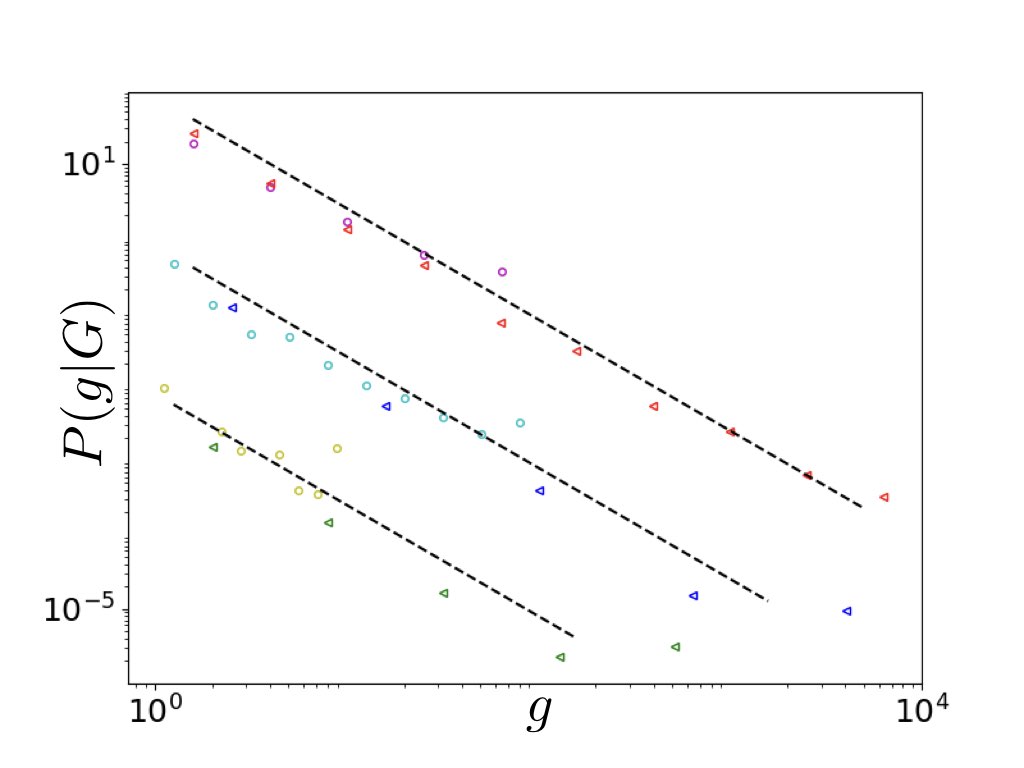}
	\caption{\textbf{Left:} Distribution of cluster sizes $S_c$ conditioned by the total size $S$ of the avalanche to which the clusters belong. Two values of $S$ are compared~: $S=10^3$ (circles) and $S=10^5$ (triangles). The conditional distribution $P(S_c|S)$ is a power law cut at $S$ with the same exponent $\tau_c$ as $P(S_c)$.
	\textbf{Middle :} Conditional distribution $P(\ell_c|\ell)$ for $\ell=10^2$ (circles) and $\ell=10^4$ (triangles). The conditional distribution $P(\ell_c|\ell)$ is a power law cut at $\ell$ with the same exponent as $P(\ell)$.
	The dashed lines are fits using the exponents $\tau_c$ (left panel) and $\kappa_c$ (middle) determined in figure~\ref{fig: P(S_c) and P(ell_c)} and listed in table~\ref{tab: measurement with errors}.
	\textbf{Parameters values :} $\alpha=0.5$, $m^2=10^{-3}$, $L=2^{17}$ ; 
	$\alpha=0.75$, $m^2=3.10^{-4}$, $L=2^{17}$ ; $\alpha=1$, $m^2=3.10^{-4}$, $L=2^{18}$ ; 
	$\alpha=1.5$, $m^2=3.10^{-5}$, $L=2^{17}$.
	\textbf{Right :} Conditional distribution $P(g|G)$. For each $\alpha$ two values of $G$ are compared~: $G=10^2$ (circles) and $G=10^4$ (triangles) for $\alpha=0.5$ and $\alpha=1$ and $G=10$ (circles) and $G=10^3$ (triangles) for $\alpha=1.5$. The conditional distribution $P(g|G)$ is a power law cut at $G$ that is fitted with the same exponent $\kappa_g=3/2$ as $P(g)$.
	\textbf{Parameters values :} $\alpha=0.5$, $m^2=0.01$, $L=2^{17}$ ; 
	$\alpha=1$, $m^2=0.05$, $L=2^{17}$ ; 
	$\alpha=1.5$, $m^2=0.03$, $L=2^{17}$.
	Data have been shifted vertically to enhance visibility.
	\label{fig: P(S_c|S) and P(ell_c|ell)}} 
	\end{adjustwidth}
\end{figure}

\paragraph{\textbf{Relation between $\tau_c$ and $\gamma_S$}}

An important assumption of the derivation proposed in Ref.~\cite{laurson2010} is that the cluster size
scales as $S_c \sim S/\langle N_c \rangle_S$. This means that for a given avalanche the clusters have similar sizes. As a consequence one could set $P(S_c)dS_c  = P(S) dS$ which yields 
$\tau_c = (\tau - \gamma_S) / (1-\gamma_S)$.
We argued that the latter relation is wrong because the clusters in a given avalanche display very different sizes, from very small up to the avalanche size. 
In figure~\ref{fig: P(S_c|S) and P(ell_c|ell)} we show the distribution $P(S_c|S)$ of cluster sizes conditioned by the total size $S$ of the avalanche they belong to. $P(S_c|S)$ displays a power law for 
$S_c$ up to $S$ that we fit with the same exponent $\tau_c$ as the unconditioned distribution $P(S_c)$.
Note that the normalization constant does not depend on $S$ (as one can see in figure~\ref{fig: P(S_c|S) and P(ell_c|ell)}) but only on a small scale cutoff $S_0$. 
Indeed we have~:
\begin{equation}
\int_{S_0}^{S} S_c^{-\tau_c} dS_c = \frac{S_0^{1-\tau_c} - S^{1-\tau_c}}{\tau_c-1} 
	\simeq \frac{S_0^{1-\tau_c}}{\tau_c-1}   \, . \notag
\end{equation}
Therefore the scaling for the mean size of the clusters belonging to avalanches of size $S$ is~:
\begin{equation}
\langle S_c \rangle_S \sim \int_{S_0}^{S} S_c^{1-\tau_c} dS_c \sim S^{2-\tau_c} \, .
\end{equation}
In consequence, using the equality~\eqref{eq: equality S_c average} we have $\langle N_c \rangle_S = S/\langle S_c \rangle_S \sim S^{\tau_c-1}$
which implies the relation~:
\begin{equation}
\gamma_S = \tau_c - 1 \, . \label{eq: tau_c = gamma_s+1}
\end{equation}

\paragraph{\textbf{Relation between $\kappa_c$ and $\gamma_{\ell}$}}
A similar argument holds for the scaling of the cluster extensions. The middle panel of figure~\ref{fig: P(S_c|S) and P(ell_c|ell)} shows that the extensions of clusters belonging to avalanches of linear size $\ell$ are power law distributed, up to the size $\ell$, with the same exponent $\kappa_c$ as the unconditioned distribution $P(\ell_c)$ This implies that
$\langle \ell_c \rangle_\ell \sim \ell^{2-\kappa_c}$ and thus~:
\begin{align}
\langle N_c \rangle_{\ell} &\sim \ell^{\gamma_{\ell}} \quad \text{with} \label{eq: Nc sim ell gamma_ell}\\
\gamma_{\ell} &= \kappa_c-1 \, , \label{eq: gamma_ell = kappa_c-1}
\end{align}
where we have made use of the equality~\eqref{eq: equality ell_c}.
We perform fits of the relation~\eqref{eq: Nc sim ell gamma_ell} in the right panel of figure~\ref{fig: P(N_c|S) and Nc vs ell} which yield the values of $\gamma_{\ell}$ listed in table~\ref{tab: measurement with errors}.

\paragraph{\textbf{Conditional distribution $P(g|G)$}}
Finally the right  panel of figure~\ref{fig: P(S_c|S) and P(ell_c|ell)} shows the distribution $P(g|G)$ of gaps $g$ conditioned by the total gap length $G$ of the avalanche they belong to. It also displays a power law, up to $G$, with the same exponent 
$\kappa_g=3/2$ as the unconditioned distribution $P(g)$. We use this assumption to derive the scaling relation for the avalanche diameter in the main text.

\begin{figure}
	\begin{adjustwidth}{-1cm}{-1cm}
	\centering
	\includegraphics[scale=0.25]{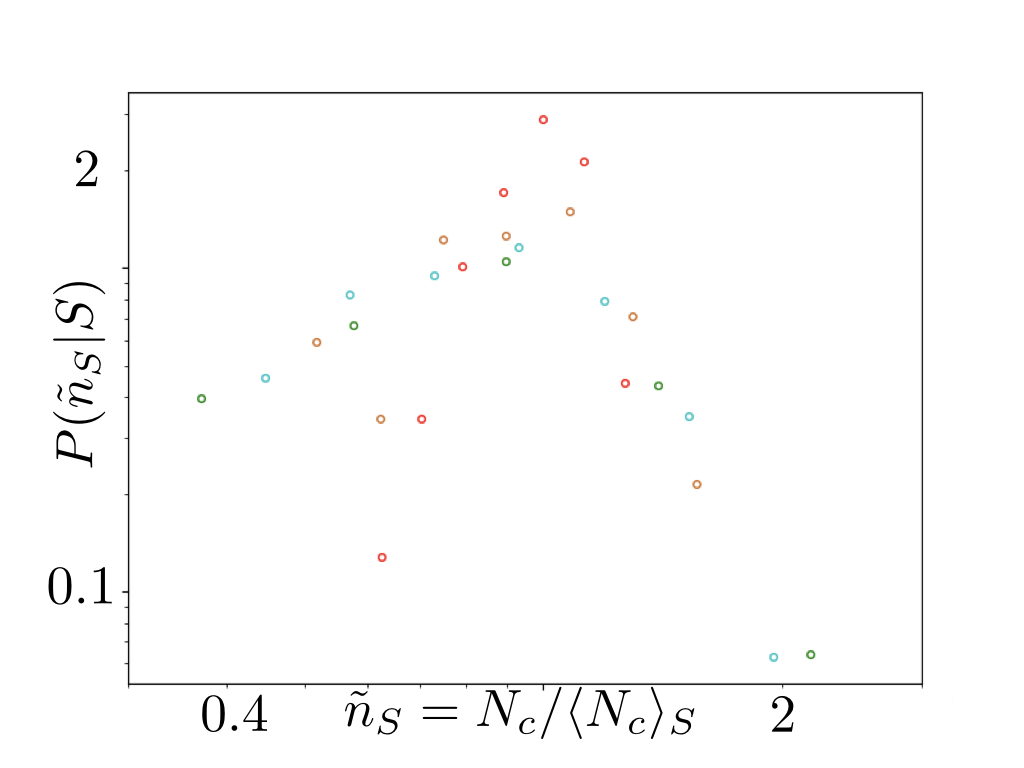}\includegraphics[scale=0.25]{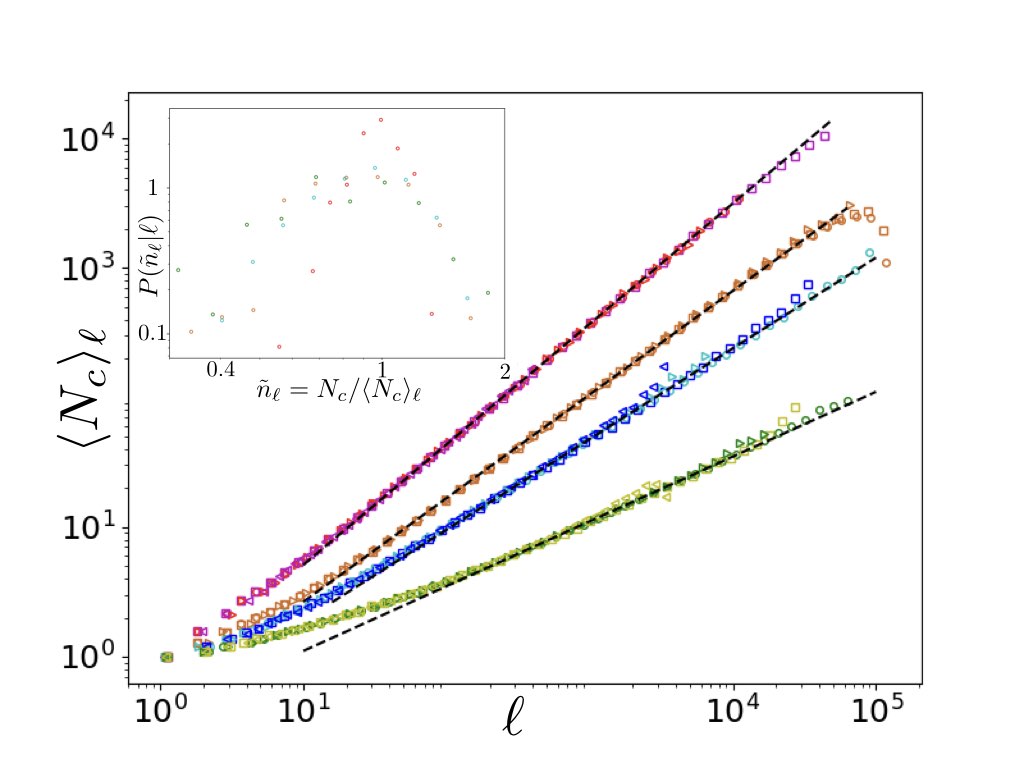}
	\caption{\textbf{Left :} Probability of $\tilde{n}_{S} = N_c/\langle N_c \rangle_{S}$ conditioned by the avalanche size $S$. The conditional probability has been computed in small intervals centered around $\ell=10^3$, $10^3$, $10^4$ and $10^5$ for $\alpha=0.5$, 0.75, 1 and 1.5 respectively.
	\textbf{Right :} $\langle N_c \rangle_{\ell}$ versus $\ell$. The dashed lines correspond to the fit $\langle N_c \rangle_{\ell} \sim \ell^{\gamma_{\ell}}$ with the values of $\gamma_{\ell}$ given in table~\ref{tab: measurement with errors}.
	\textbf{Inset :} Probability of $\tilde{n}_{\ell} = N_c/\langle N_c \rangle_{\ell}$ conditioned by the avalanche linear size $\ell$. The conditional probability has been computed in small intervals centered around $\ell=10^2$, $10^2$, $10^3$ and $10^4$ for $\alpha=0.5$, 0.75, 1 and 1.5 respectively.
	\label{fig: P(N_c|S) and Nc vs ell}} 
	\end{adjustwidth}
\end{figure}

\paragraph{\textbf{Justification for using $P(S)dS = P(N_c)dN_c = P(\ell)d\ell$}}

To derive our scaling relations for $\tau_c$ and $\kappa_c$ we combine our numerical observation that the distribution of $N_c$ is originated by a BGW process with the assumption $P(S)dS=P(N_c)dN_c=P(\ell)d\ell$. 
The latter requires that the conditional distributions $P(N_c|S)$ and $P(N_c|\ell)$ are peaked. We verify this condition in figure~\ref{fig: P(N_c|S) and Nc vs ell}. 
Our assumption gives the relations~:
\begin{align}
\gamma_{S}&=2(\tau-1) \, , \quad \gamma_{\ell}=2(\kappa-1)\, ,\label{eq: gamma_S = 2 tau - 1} 
\end{align}
that we combine with equations~\eqref{eq: tau_c = gamma_s+1} and \eqref{eq: gamma_ell = kappa_c-1} to obtain the scaling relations of the main text~:
\begin{align}
\tau_c &=2\tau-1 \, , \quad  \kappa_c =2\kappa-1 \, .\label{eq: tau_c = 2 tau - 1} 
\end{align}

\end{widetext}
\end{document}